\documentclass[twocolumn,aps,showpacs,preprintnumbers,amsmath,amssymb]{revtex4}
\usepackage{graphicx}

\begin{document}

\title{Quantum Properties of a Nanomechanical Oscillator}

\author{Aziz Kolkiran}
\author{G. S. Agarwal}
\affiliation{Department of Physics, Oklahoma State University,
Stillwater, OK - 74078, USA}
\date{submitted to {\bf Phys. Rev. B} on August 28, 2006}

\begin{abstract}
We study the quantum properties of a nanomechanical oscillator via
the squeezing of the oscillator amplitude. The static longitudinal
compressive force $F_0$ close to a critical value at the Euler
buckling instability leads to an anharmonic term in the
Hamiltonian and thus the squeezing properties of the
nanomechanical oscillator are to be obtained from the Hamiltonian
of the form  $H= a^{\dag}a+\beta (a^{\dag}+a)^4/4$. This
Hamiltonian has no exact solution unlike the other known models of
nonlinear interactions
 of the forms
$a^{\dag 2}a^2$, $(a^{\dag}a)^2$ and
$a^{\dag4}+a^4-(a^{\dag2}a^2+a^2a^{\dag2})$ previously employed in
quantum optics to study squeezing. Here we solve the Schr\"odinger
equation numerically and show that in-phase quadrature gets
squeezed for both ground state and coherent states. The squeezing
can be controlled by bringing $F_0$ close to or far from the
critical value $F_c$. We further study the effect of the
transverse driving force on the squeezing in nanomechanical
oscillator.
\end{abstract}
 \maketitle

\section{Introduction}
There is currently a wide effort to observe quantum behavior in
nanoscale devices \cite{Gaidarzhy 2005, Cho 2003, Blencowe 2004,
LaHaye 2004}. In the limit of high resonator frequency with high
mechanical quality factors and long coherence lifetimes, the
nanomechanical oscillator (NMO) phonons will be analogous to
photons in an electromagnetic cavity. With current technology it
is possible to reach resonator frequency of GHz order \cite{Huang
2003}. At a temperature of around 50mK, one can principally
prepare the resonator into the ground state. These sub-Kelvin
temperatures are well within the range of todays dilution
refrigerators. However cooling the resonator down to these
temperatures requires some other techniques \cite{Wilson-Rae 2004,
Zhang 2005}.
\begin{figure}
  \centering
  \scalebox{1.25}{\includegraphics{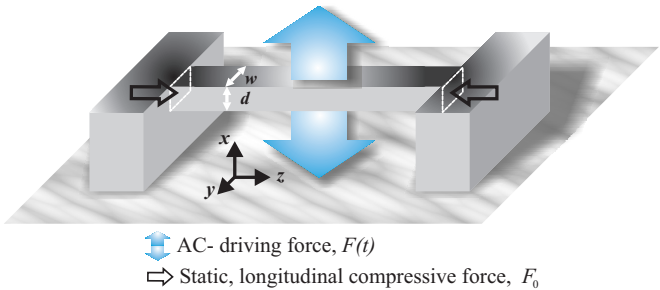}}
  \caption{(Color online) Schematic diagram of the freely suspended nanomechanical beam of total length $L$, width $w$ and thickness $d$.
  The beam is clamped at both ends. A static, mechanical force $F_0$ compresses the beam in longitudinal
  direction controlling the nonlinearity. An ac-driving force can be used to excite the beam to transverse vibrations.}
  \label{Fig01}
\end{figure}

With the assumption that quantum mechanics should apply to these
mesoscopic systems, variety of methods and techniques have been
proposed to observe some quantum optical effects including
solid-state laser cooling \cite{Wilson-Rae 2004, Zhang 2005},
quantum nondemolition measurement \cite{Santamore 2004, Ruskov
2005}, phonon lasing \cite{Bargatin 2003} and squeezed state
generation \cite{Blencowe 2000, Garrett 1997}. There are also
proposals analyzing macroscopic quantum tunneling \cite{Carr
2001}, resonant multi-phonon excitations \cite{Peano 2004} and a
variety of methods to entangle mechanical resonators with other
quantum systems \cite{Armour 2002, Cleland 2004, Geller 2005, Bose
2006}.

The next question is what could be the best way to study quantum
properties of a NMO. In line with the work in quantum optics on
squeezing, we can consider studying the squeezed states of the
NMO. One proposal considers modulating the spring constant to
produce squeezing \cite{Blencowe 2000} as it is known from the
earlier work \cite{Agarwal 1991} that any modulation of the
frequency of the oscillator can result in squeezing. Here we adopt
a different model. We consider the situation shown schematically
in the Fig. \ref{Fig01}. We show various forces acting on a
nanobeam structure which is clamped at both ends vibrating in the
transverse direction. There is a static mechanical force, $F_0$,
acting in the longitudinal direction and an ac-driving force to
excite vibrations in the transverse direction. The longitudinal
force $F_0$ which is close to a critical value at the Euler
instability gives rise to an additional term in the potential
energy which is quartic in the fundamental mode amplitude $x$. The
effective Hamiltonian that describes the system would be in the
form $H=p^2/2m+m\omega^2 x^2/2+\tilde{\beta} x^4/4$. The
derivation of this nonlinearity in $x$ is given in the next
section.
 Unlike the previous work on squeezing \cite{Blencowe
2000} in a NMO we would consider the effect of the nonlinearity in
$x$. Note that the nonlinearity can be switched on and off by
controlling $F_0$. We would thus study the quantized behavior of a
NMO subject to the force $F_0$.

The organization of the paper is as follows. The model is
described in section \ref{model} and the effective Hamiltonian is
derived briefly by referring to the previous works for the doubly
clamped elastic rectangular beam. We discuss the previous works on
the squeezing in nonlinear oscillators in section \ref{previous
work}. Then, we study the quantum dynamics and analyze the
squeezing properties in section \ref{dynamics}. The conclusion and
future perspective are given in section \ref{conclusion}.

\section{ The Model} \label{model}

We start with an elastic rectangular beam of length $L$, width $w$
and thickness $d$ as shown in Fig. \ref{Fig01}. The beam is freely
suspended and clamped at both ends. The transverse motion in the
direction of $d$ is allowed. The dimensions are such that ($L\gg
w>d$) there is no appreciable vibrations in other directions. A
static mechanical force $F_0$ acts on the beam in the longitudinal
direction ($F_0>0$ for compression). An ac-driving field,
$F(t)=\widetilde{f}\cos(\omega_{ex}t)$ can also be added to excite
the vibrations. The dynamics of the beam can be completely
described by the transverse deflection $\phi(s)$ parametrized by
the arclength $s\in[0,L]$ in a classical picture. Assuming single
transverse degree of freedom for simplicity the nonlinear
Lagrangian of the system, for arbitrary strong deflections
$\phi(s)$ is then \cite{Poston 1978, Werner 2004},
\begin{widetext}
\begin{equation}\label{lagrangian}
    \mathcal{L}(\phi,\dot{\phi},t)=\int_0^L
    ds\left[\frac{\rho}{2}\dot{\phi}^2-\frac{\mu}{2}\frac{\phi''^2}{(1-\phi'^2)}-F_0(\sqrt{1-\phi'^2}-1)+F(t)\phi\right].
\end{equation}
\end{widetext}
Here $\rho=m/L$ is the mass density, $\mu=EI$ is the product of
the elasticity modulus $E$ and the moment of inertia $I$. In
$\phi'$, prime denotes partial derivative with respect to $s$,
i.e. $\partial\phi/\partial s$. For small oscillations
$|\phi'(s)|\ll1$, the Lagrangian is quadratic and it leads to the
linear equation of motion
\begin{equation}\label{eqofmotion1}
    \rho\ddot{\phi}+\mu\phi''''+F_0\phi''=0.
\end{equation}
The equation of motion can be separated and transformed into an
eigenvalue problem with boundary conditions applied to the
endpoints. One can write the general solution as a superposition
$\phi(s,t)=\sum_n \phi_n(s,t)=\sum_n\mathcal{A}_n(t) g_n(s)$,
where $g_n(s)$ are the normal modes which follow as solution of
the characteristic equation and $n=1$ is called the fundamental
mode. For the doubly clamped nanobeam, we have $\phi(0)=\phi(L)=0$
and $\phi'(0)=\phi'(L)=0$. The expressions for $\phi_n$'s are then
given by a superposition of trigonometric and hyperbolic
functions, and the eigenfrequencies
$\omega_n$'s are the solutions of transcendental equations. 
For the doubly clamped boundary conditions, when $F_0$ is close to
the critical value $F_c=\mu(\pi/L)^2$, one can get the simplified
fundamental frequency $\omega_1(F_0\rightarrow
F_c)=\sqrt{\epsilon}\omega_0$ where
$\omega_0=(2\pi^2/3)\sqrt{E/\rho}(d/L^2)$ is the fundamental
frequency of the relaxed beam($F_0=0$). The parameter
$\epsilon=(F_c-F_0)/F_c$ is called the distance to the critical
force and the system reaches to the well known Euler instability
when $\epsilon\ll1$. The dynamics at low energies close to the
Euler instability will be dominated by the fundamental mode alone.
The fundamental mode $g_1(s)$ can also be expanded close to the
Euler instability in zeroth order in $\epsilon$
\begin{equation}\label{fundamental mode}
g_1(s)\simeq\sin^2\left(\frac{\pi s}{L}\right).
\end{equation}
 Since the
fundamental frequency $\omega_1$ vanishes at the critical value
$F_c$, one has to include the contributions beyond the quadratic
terms $\phi'^2$ and $\phi''^2$ in the Lagrangian. The next higher
order terms, $-(\mu/2)\phi''^2\phi'^2$ and $(F_0/4)\phi'^4$, are
quartic in the Lagrangian. Inserting the normal mode expansion
$\phi(s,t)=\sum_n\mathcal{A}_n(t) g_n(s)$ in the Lagrangian and
assuming that the fundamental mode $n=1$ dominates the dynamics
(by neglecting the higher modes $n=2,3,\ldots$) one can quantize
the theory by introducing the canonically conjugate momentum
$p\equiv -i\hbar\partial/\partial\mathcal{A}_1$ with the
``coordinate'' $x\equiv \mathcal{A}_1$. Note that when the driving
frequency is close to the fundamental frequency of the beam, the
fundamental mode will dominate also in absence of a static
longitudinal compression force $F_0$. However, a compression force
close to a critical value helps to enhance the nonlinear effects
which are of the importance of this paper.

By using the above definition of coordinate and the conjugate
momentum, an effective quantum mechanical time-dependent
Hamiltonian results describing the dynamics of a single quantum
particle
\begin{equation}\label{Hamiltonian}
    \widetilde{H}(t)=\frac{p^2}{2m^*}+\frac{m^*\omega_1^2}{2}x^2+\frac{\widetilde{\beta}}{4}x^4+
    xF(t),
\end{equation}
with the effective mass $m^*=3\rho L/8$, the fundamental frequency
$\omega_1=\sqrt{\epsilon}(2\pi^2/3)\sqrt{E/\rho}(d/L^2)$ and the
nonlinearity parameter
$\widetilde{\beta}=(\pi/L)^4F_cL(1+3\epsilon)$ \cite{Peano 2006}.
Now, Eq. (\ref{Hamiltonian}) can be put in a second-quantized form
by replacing $x$ and $p$ with the creation and annihilation
operators $a^{\dag}$ and $a$,
\begin{eqnarray}\label{operators}
   x=\sqrt{\frac{\hbar}{m^*\omega_1}}\frac{1}{\sqrt{2}}(a^{\dag}+a)\\
   p=\sqrt{m^*\hbar\omega_1}\frac{i}{\sqrt{2}}(a^{\dag}-a).
\end{eqnarray}
Upon scaling the Hamiltonian by $\hbar\omega_1$ we obtain the
dimensionless form,
\begin{widetext}
\begin{equation}\label{Ham2}
    H(t)=a^{\dag}a+\frac{1}{2}+\beta\left(\frac{a^{\dag}+a}{\sqrt{2}}\right)^4+
    f\cos(\omega_{ex}t)\left(\frac{a^{\dag}+a}{\sqrt{2}}\right),
\end{equation}
\end{widetext}
with the redefined dimensionless parameters,
\begin{equation}\label{parameters}
    \beta=\frac{\widetilde{\beta}x_0^4}{4\hbar\omega_1},\hspace{0.5cm}
    f=\frac{\widetilde{f}x_0}{\hbar\omega_1},
\end{equation}
where $x_0=\sqrt{\hbar/m^*\omega_1}$. One can obtain an expression
for $\beta$ that depends on the dimensions and the material
properties of the beam. By substituting the parameters in Eq.
(\ref{parameters}) one finds
\begin{equation}\label{beta}
   \beta=\frac{\hbar}{2}\frac{1}{L\,w\,d^2}\frac{1}{\sqrt{\rho
   E}}\frac{1+3\epsilon}{\epsilon^{3/2}}.
\end{equation}
Note that the equation (\ref{beta}) is valid for $\epsilon\ll 1$
and $\beta$ can be controlled by fine tuning the distance
parameter $\epsilon$ at this regime. Table \ref{table1} lists the
range of the nonlinearity parameter $\beta$ as well as the relaxed
fundamental frequencies and the critical compressions for three
different sizes of Si nanobeams. Note that one should have
extremely precise control over $\epsilon$ to increase the
nonlinearity. In light of the measurements done in the experiment
\cite{Babic 2003} $\epsilon$ was found to be of the size
$\thickapprox 10^{-5}$ for a 100 nm length carbon nanotube.
\begin{table}[h]
  \caption{\label{table1}Table shows calculated $\beta$ values, the nonlinearity parameter, for two different beam size.
   For each size, the relaxed fundamental mode frequency $\omega_0$ and the critical force $F_c$ defined in the text, are also shown. The parameter
   $\epsilon$ shows the distance to the critical force near the Euler instability. }
  \begin{center}
  \begin{tabular}{l|l|lll} \hline\hline
 \multicolumn{5}{c}{Si bar,  \hspace{0.5cm}$\rho=2330\,\mathrm{kg/m^3}$, \hspace{0.5cm}$E=169\,\mathrm{GPa}$}
 \\\hline\hline
 $L=200$ nm   &   $\,\,\omega_0/2\pi=1.12$ GHz  &    $\,\,\epsilon=10^{-6}$    &  $\Longrightarrow$ & $\,\,\beta=0.05$  \\
 $d=5$ nm     &   $\,\,F_c=4.34$ nN             &    $\,\,\epsilon=10^{-7}$    &  $\Longrightarrow$ & $\,\,\beta=1.68$  \\
 $w=10$ nm    &                                 &                              &                    &                   \\
 \hline\hline
 $L=400$ nm   &   $\,\,\omega_0/2\pi=557$ MHz   &    $\,\,\epsilon=10^{-7}$    &  $\Longrightarrow$ & $\,\,\beta=0.11$  \\
 $d=10$ nm    &   $\,\,F_c=17.3$ nN             &                              &                    &                   \\
 $w=20$ nm    &                                 &                              &                    &                   \\
 \hline\hline
  \end{tabular}
  \end{center}
  \end{table}
\section{Previous works on the squeezing in nonlinear
oscillators}\label{previous work}

The squeezing produced by the nonlinearity has been investigated
in the past by using a number of approximations however none of
these are suitable for the problem of the NAMO. Milburn dropped
all phase sensitive terms from (\ref{Ham2}) and studied
\cite{Milburn 1986} the simplified Hamiltonian,
\begin{equation}\label{Milburn hamiltonian}
   H=\hbar\omega [(1+\beta)a^{\dag}a+\beta (a^{\dag}a)^2].
\end{equation}
By solving exactly the phase space distribution function  he
showed that squeezing can be obtained for a coherent state of
amplitude $\alpha=0.5$ for very short times. Buzek \cite{Buzek
1989} and Tanas \cite{Tanas 1989} studied  Hamiltonian models of
the form,
\begin{eqnarray}
   H=\hbar\omega a^{\dag}a+\frac{1}{2}\hbar\beta a^{\dag
   2}a^2,\label{buzek hamiltonian}\\
   H=\hbar\omega a^{\dag}a+\frac{1}{2}\hbar\beta (a^{\dag}a)^2.\label{tanas hamiltonian}
\end{eqnarray}
They solved the Heisenberg equations of motion exactly and showed
periodic squeezing for coherent states in both quadratures. Tanas
\cite{Tanas 1989} showed also that maximum squeezing can be
obtained in the limit of large mean number ($|\alpha|^2\gg1$) and
small times ($t\ll 1$). No squeezing is allowed for the vacuum
(ground) state in the above models.

In this paper, we calculate the squeezing in an anharmonic
oscillator for the anharmonicity quartic in $x$ which is the
amplitude of the fundamental mode of the oscillation. One could
write $x^4$ in terms of creation and annihilation operators as
follows:
\begin{equation}\label{x4}
    x^4\propto(a^{\dag}+a)^4=a^{\dag 4}+a^4+f(a^{\dag},a)
\end{equation}
where $f(a^{\dag},a)$ is a polynomial in $a$ and $a^{\dag}$ of
order four which is given by
\begin{eqnarray}\label{f4}
    f(a^{\dag},a)=&6&\!\!\!a^{\dag2}a^2+4(a^{\dag2}a^{\dag}a+a^{\dag}aa^2)+6(a^{\dag
    2}+a^2)\nonumber\\
    &+&12a^{\dag}a+3.
\end{eqnarray}
The Hamiltonian containing the second and third terms given in Eq.
(\ref{x4}) give rise to two-photon (or phonons in quantum
mechanical descriptions of solid systems) transitions. It is known
that two-photon transitions are necessary for producing squeezed
states and thus the terms $a^{\dag}aa^2$, $a^2$ and their
hermitian conjugate should be important. In the literature,
multi-photon processes have also been analyzed to study normal and
higher order squeezing in the limit of small times \cite{Buzek
1990, Braunstein 1987, Hillery 1984}. In relation to this paper,
Tombesi and Mecozzi \cite{Tombesi 1988} studied the harmonic
oscillator model which has four-photon transitions in the
interaction term,
\begin{equation}\label{f4_2}
     H_I=\beta[a^{\dag4}+a^4-(a^{\dag2}a^2+a^2a^{\dag2})].
\end{equation}
This model was solved exactly. They showed that significant amount
of normal and higher order squeezing is possible for initial
coherent states of amplitude $|\alpha|>1$ with certain phases and
for short times. No squeezing is allowed for the vacuum and
fluctuations diverge as time grows.
\begin{figure*}
  \centering
  \scalebox{0.35}{\includegraphics{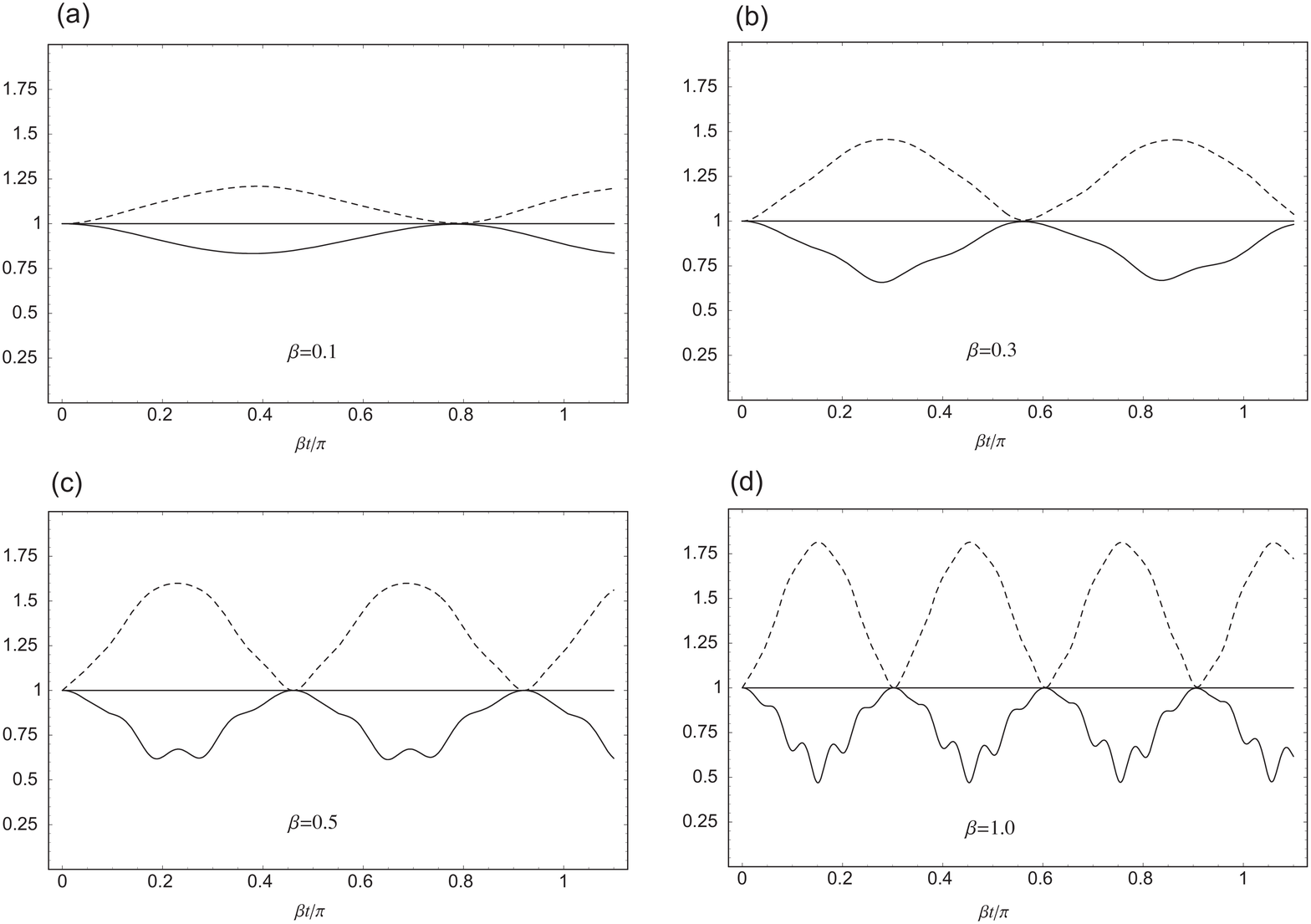}}
  \caption{Squeezing of the ground state in the anharmonic oscillator model given by Eq. (\ref{Ham2})
  changing from weak to strong nonlinearity $\beta$; (a) $\beta=0.1$, (b) $\beta=0.3$, (c)
  $\beta=0.5$, (d) $\beta=1.0$. Solid line, $S_x(t)$, normally ordered normalized fluctuations in $x$ and dashed line,
$S_p(t)$, normally ordered normalized fluctuations in $p$ as
   given by Eqs. (\ref{xsqueezing}) and (\ref{psqueezing}) respectively.}
  \label{Fig02}
\end{figure*}
\begin{figure*}
  \centering
  \scalebox{0.35}{\includegraphics{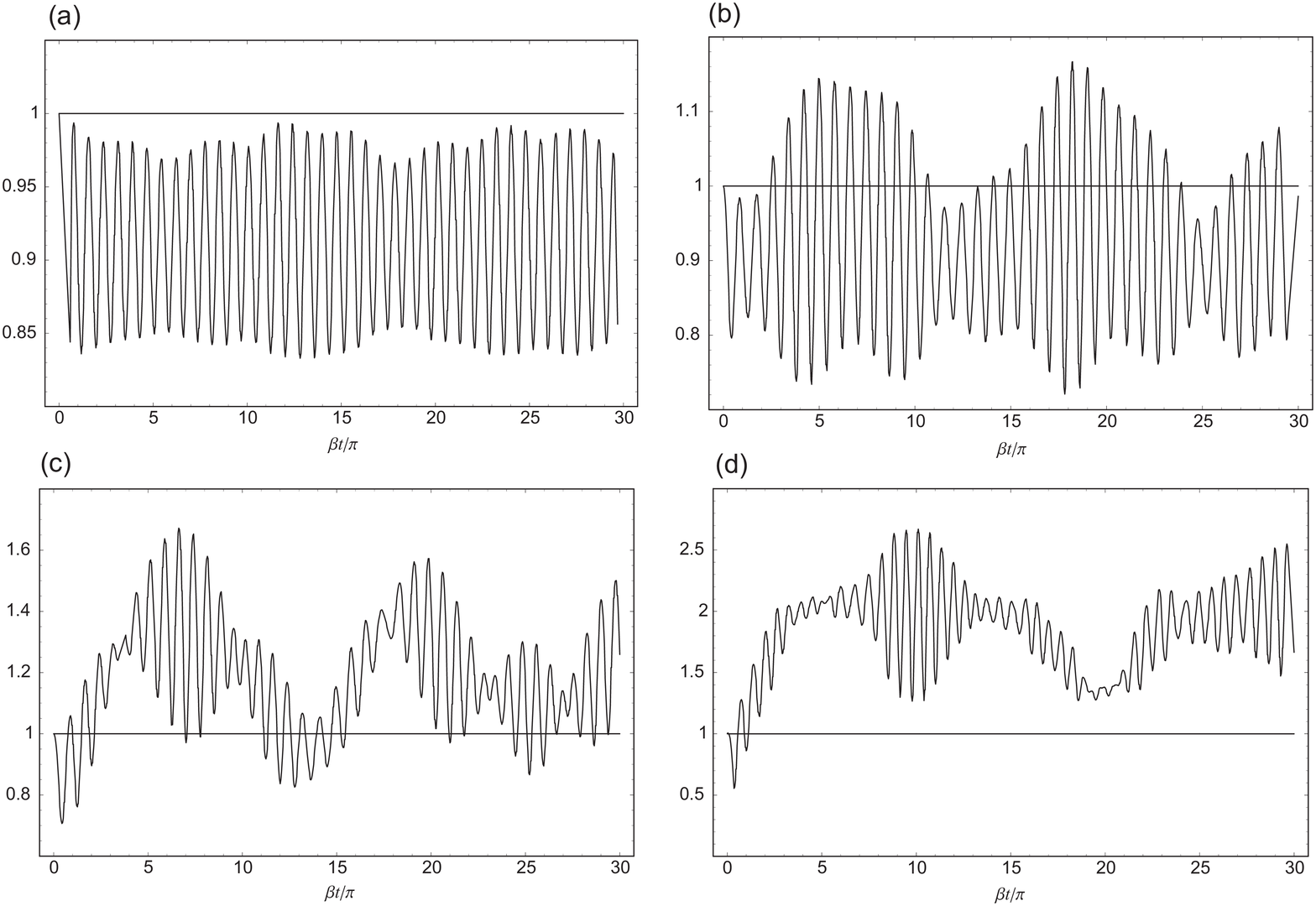}}
  \caption{The squeezing $S_x(t)$ for the coherent state of amplitudes,
  (a) $|\alpha|=0.1$, (b) $|\alpha|=0.5$, (c) $|\alpha|=1.0$, (d) $|\alpha|=2.0$.
  The phase of $\alpha$ is equal to $\pi/2$. The nonlinearity parameter $\beta$ is taken as 0.1.}
  \label{Fig03}
\end{figure*}

In the description of NMO systems the anharmonic Hamiltonian
models given in Eqs. (\ref{Milburn hamiltonian}), (\ref{buzek
hamiltonian}), (\ref{tanas hamiltonian}) and (\ref{f4}) are no
good for vacuum (ground state) squeezing. Moreover, the nonlinearity $\lambda$
cannot be controlled externally since it is an intrinsic property
of the medium. To observe the nonclassical (quantum) properties of
a mesoscopic system in general, the control of the parameter that
gives rise to the nonclassical behavior would be crucial for an
experimentalist.

The harmonic oscillator having the nonlinearity of Eq. (\ref{x4})
that we shall work in the next sections, gives important squeezing
in the in-phase quadrature for both ground state and coherent
states. Furthermore, it will be shown in section III that the
physical model of the NMO allows one to control the nonlinearity
by the application of a static external force. The numerical
solution of the Hamiltonian shows that the ground state squeezing
displays periodicity and it stays squeezed for the whole cycle of
the period. In fact, the ground state squeezing is important for
the mesoscopic resonators because bringing the harmonic oscillator
representing the nanomechanical system to its vibrational ground
state is a necessary prerequisite for quantum state engineering.
The effect of driving term is also examined.

\section{Quantum dynamics and Squeezing}\label{dynamics}

We first consider the case in which there is no driving. The
Hamiltonian
\begin{equation}\label{Ham3}
    H=a^{\dag}a+\frac{1}{2}+\beta\left(\frac{a^{\dag}+a}{\sqrt{2}}\right)^4,
\end{equation}
has no analytical solution. For the numerical calculations, we
employ the split-operator method \cite{Fleck 1976} for the time
propagation of the initial state. In this method, one can split
the propagator on a time step $\Delta t$ as
\begin{eqnarray}\label{split-operator}
   U(t+\Delta t)&=&e^{-i(H_0+V)\Delta t}\\
   &=&e^{-iH_0\Delta t/2}e^{-iV\Delta t}e^{-iH_0\Delta t/2}+O[(\Delta
   t)^3]\nonumber,
\end{eqnarray}
where $H_0=a^{\dag}a+1/2$ and $V=\beta (a^{\dag}+a)^4/4$. That
means splitting the exponential of the operators which are not
commuting is accurate to second order in the time step $\Delta t$.
Therefore one can make the calculation as accurate as possible by
taking the time step sufficiently small. Now we are ready to
analyze the dynamics of the oscillations of the fundamental mode.
We take the ground state of the Hamiltonian $H_0$ as our initial
state. Then we analyze the dynamics by the application of
compressive force $F_0$ suddenly (in a time interval much faster
than the oscillator's frequency) in the regime near to the Euler
instability where the nonlinearity parameter $\beta$  can go from
negligibly small values (where the dynamics is largely governed by
) to appreciable values. One can see how the nonlinearity
parameter $\beta$  changes by tuning the distance parameter
$\epsilon$ in the table \ref{table1}. We examine the dynamics by
calculating the time evolution of the width of initial wave
function in the ``coordinate"-space, i.e. $x$-space (we take the
amplitude of the transverse deflection of the NMO as our
coordinate as explained in section II.) and the corresponding
momentum-space, i.e. $p$-space. Next, we define the squeezing
factors
\begin{eqnarray}
   S_x(t)=\frac{\Delta x(t)}{\Delta
   x_0},\label{xsqueezing}\\
   S_p(t)=\frac{\Delta p(t)}{\Delta
   p_0},\label{psqueezing}
\end{eqnarray}
where $\Delta x(t)=(\langle x^2(t)\rangle-\langle
x(t)\rangle^2)^{1/2}$, and $\Delta p(t)=(\langle
p^2(t)\rangle-\langle p(t)\rangle^2)^{1/2}$. $\Delta x_0$ and
$\Delta p_0$ are the widths in coordinate and momentum space of
the initial wave function. The squeezing occurs when one of the
expressions in Eqs. \ref{xsqueezing} and \ref{psqueezing} get less
than one and it is said to be perfect when it gets zero. Fig.
\ref{Fig02} shows squeezing for the ground state for different
nonlinearity values $\beta=0.1, 0.3, 0.5$ and $1.0$.
\begin{figure}[h]
  \centering
  \scalebox{0.48}{\includegraphics{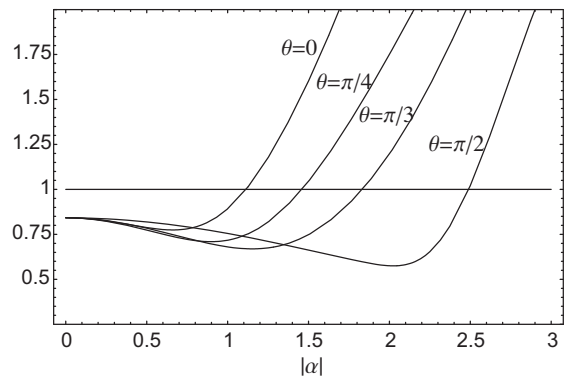}}
  \caption{Maximum squeezing in $x$ vs. coherent state amplitude $|\alpha|$. The dependence is shown for
  different values of the phase. The nonlinearity parameter $\beta$ is equal to
  0.1.}
  \label{Fig04}
\end{figure}
\begin{figure}[h]
  \centering
  \scalebox{0.45}{\includegraphics{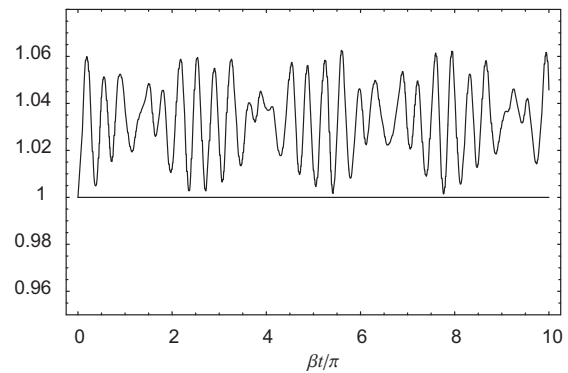}}
  \caption{Time evolution of the uncertainty product $\Delta x(t)\Delta p(t)/\Delta
x_0\Delta p_0$ for the  ground state wave function of $H_0$ under
the dynamics of the Hamiltonian $H$ given in Eq. (\ref{Ham3}).}
  \label{Fig05}
\end{figure}
\begin{figure*}
  \centering
  \scalebox{0.38}{\includegraphics{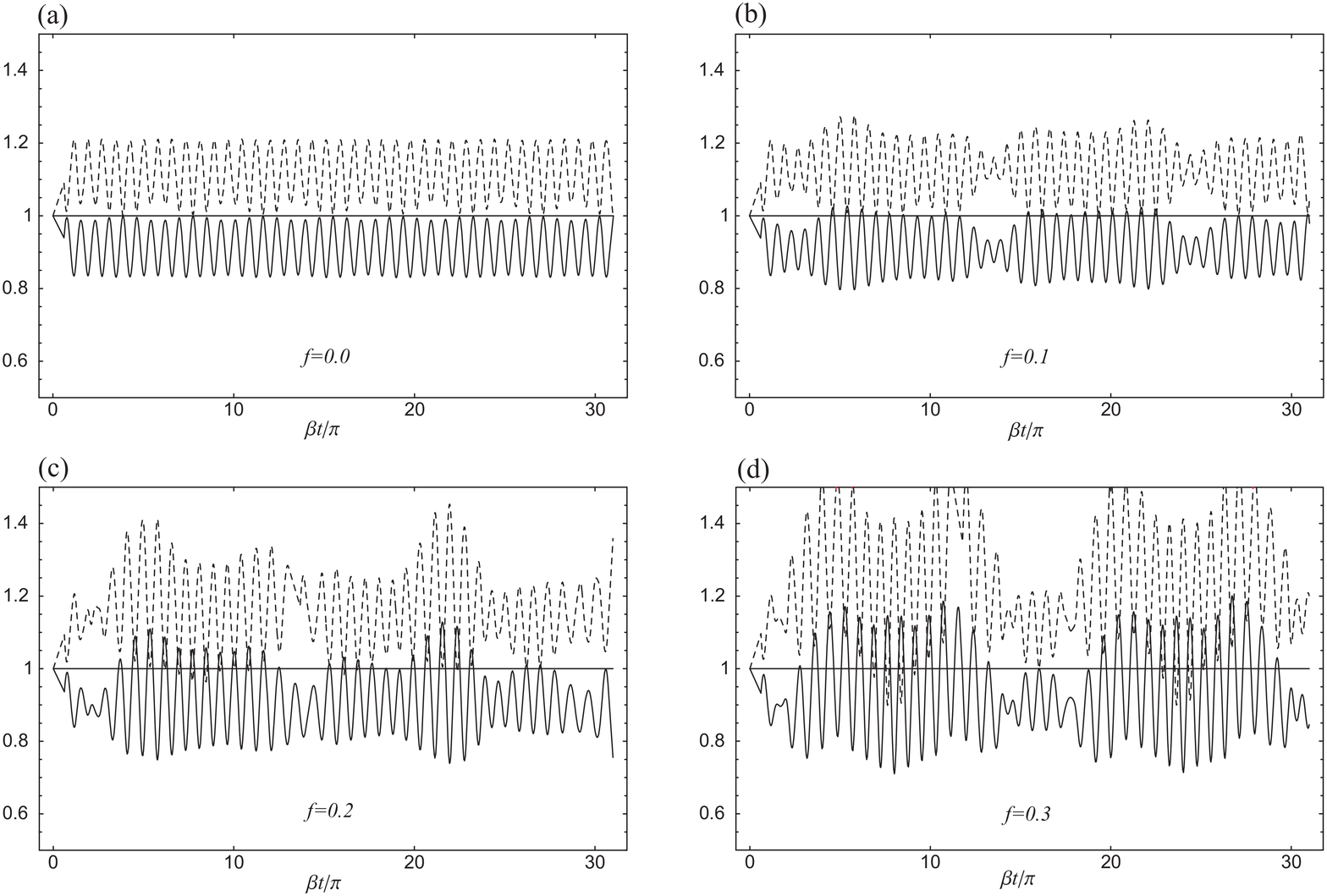}}
  \caption{The effect of the driving term on squeezing. (a) shows the squeezing without driving and (b), (c), (d)
  show the effect for the dimensionless driving parameter ($f$) values of 0.1, 0.2 and 0.3 respectively. The nonlinearity
  parameter $\beta$ is equal to 0.1. Solid line is for $S_x(t)$ and the dashed line is for $S_p(t)$.}
  \label{Fig06}
\end{figure*}

One could also start with a coherent state to check squeezing. To
prepare the coherent state, just as in the preparation of ground
state mentioned in the previous paragraph, one can first start
with the linear harmonic oscillator Hamiltonian $H_0$. Then the
coherent state can be obtained by resonantly driving (e.g. by a
harmonic electromotive force) \cite{Schwab 2002} the systems
ground state. Thereafter, by switching the nonlinearity suddenly
the dynamics of squeezing can be obtained. We analyzed squeezing
for different initial coherent state amplitudes and phases,
$\alpha=|\alpha|e^{i\theta}$, at the nonlinearity value of
$\beta=0.1$. Fig. \ref{Fig03} shows the time evolution of the
squeezing for increasing coherent state amplitude $\alpha$ at the
phase $\theta=\pi/2$. Fig. \ref{Fig04} shows the dependence of the
maximum squeezing to the phase. It can be seen from the figure
that the phase does not make much difference for low amplitudes
($|\alpha|<0.5$) but it changes squeezing behavior drastically for
amplitudes larger than $0.5$. Increasing the amplitude increases
the maximum squeezing whereas the state never becomes squeezed
after a short duration. On the other hand, low amplitudes shows
periodic squeezing all the time at a moderate value.

One can also plot the time evolution of the uncertainty product
$\Delta x(t)\Delta p(t)/\Delta x_0\Delta p_0$ for the ground state
wave function of the linear Hamiltonian $H_0$. Fig. \ref{Fig05}
shows that the oscillator recovers its minimum uncertainty
periodically and the fluctuation remains bounded.

Next we analyze the effect of driving in the dynamics of the
Hamiltonian given in Eq. (\ref{Ham2}). We calculate the propagator
again by using the split-operator method. This time, we include
the nonlinear term into $H_0$ and we take the time dependent
driving term as $V(t)$ to employ the splitting given in Eq.
(\ref{split-operator}). Fig. \ref{Fig06} shows the time evolution
of the squeezing factors given in Eqs. (\ref{xsqueezing}) and
(\ref{psqueezing}) for the in-phase and the out-of-phase
quadratures for driving parameter values of $f=0.0$, $0.1$, $0.2$
and $0.3$ at the nonlinearity $\beta=0.1$. We take the ground
state of the linear Hamiltonian as the initial state. The
frequency of the driving term, $\omega_{ex}$, is on resonance with
the frequency of the oscillator, $\omega_1$. As clearly seen, the
effect of the driving is to enhance the squeezing in $x$
periodically to a larger value. For the values of $f$ in
consideration, maximum squeezing goes from $17\%$ (for $f=0.0$) to
$30\%$ (for $f=0.3$). Increasing the driving further does not
enhance squeezing, rather it starts to enhance the fluctuations
and the squeezing cycle becomes shorter. And the driving
frequencies off the resonance does not help enhancing the
squeezing at all. The upper limit $\bar{f}_{max}$ for the regime
of validity of the fundamental mode description is given by the
first harmonic threshold, that is
$\bar{f}_{max}<\hbar\omega_2\approx 3\hbar\omega_1$ \cite{Carr
2001} which means for the dimensionless parameter, $f_{max}<3$.
\begin{figure*}
  \centering
  \scalebox{0.40}{\includegraphics{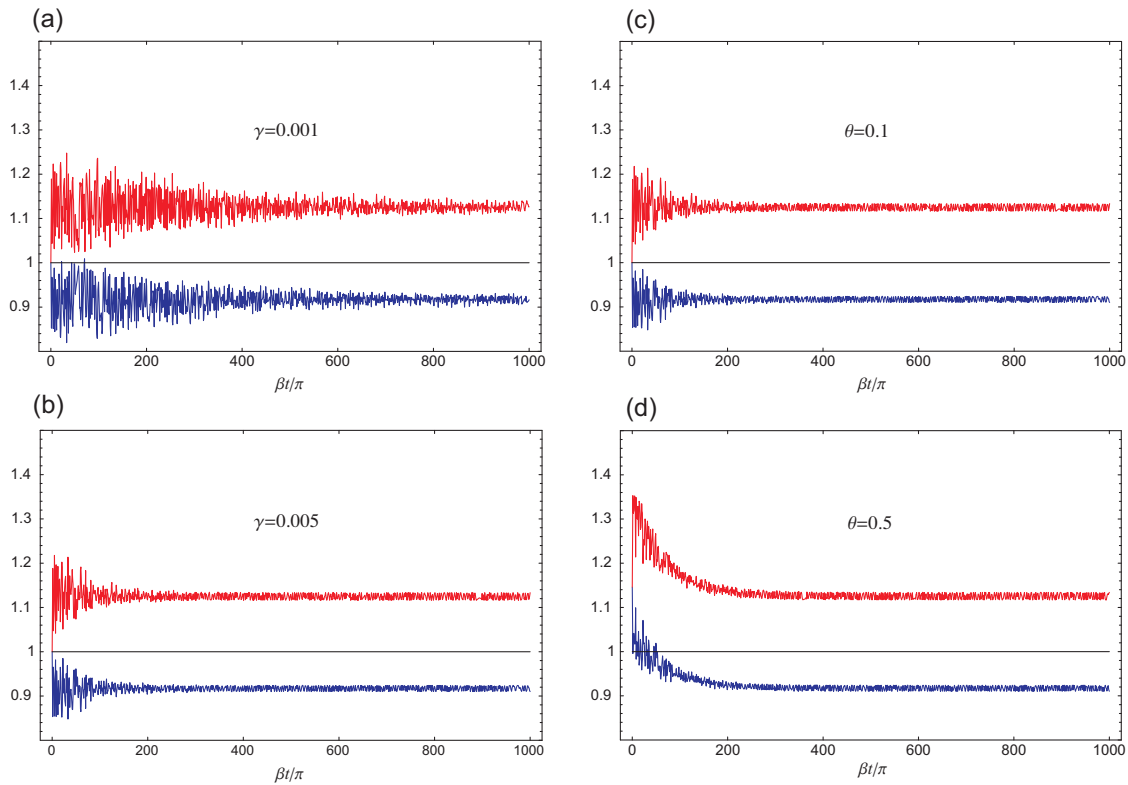}}
  \caption{(Color online) Squeezing in the driven damped anharmonic nanomechanical oscillator. The red and blue lines
  show $S_p(t)$ and $S_x(t)$ respectively. (a) and (b) show
  squeezing for the ground state with the damping constants
  $\gamma=0.001\omega_1$ and $\gamma=0.005\omega_1$ respectively. (c) and (d)
  show squeezing for the thermal state with the damping constant
  $\gamma=0.005\omega_1$ at the temperatures $kT=0.1\hbar\omega_0$ and
  $kT=0.5\hbar\omega_0$ respectively. The other parameters are kept at $\beta=0.1$ and $f=0.1$ for all.}
  \label{Fig07}
\end{figure*}

We include the damping effects in the master equation formalism of
the form
\begin{equation}\label{master equation}
\dot{\rho}=-i[H,\rho]+\gamma (a\rho
a^{\dag}-\frac{1}{2}\,a^{\dag}a\rho-\frac{1}{2}\,\rho a^{\dag}a),
\end{equation}
where $H$ is the Hamiltonian given by the Eq. (\ref{Ham2}) and
$\gamma$ is the damping constant. Here, $\gamma$ is the rate for
the oscillator to decay from the first excited state to the ground
state. The rate for the decay from level $n$ to $n-1$ is
$n\gamma$. After solving the master equation numerically we plot
the squeezing factors $S_x(t)$ and $S_p(t)$ in Figs.
\ref{Fig07}(a) and \ref{Fig07}(b) for the ground state. As seen
clearly, the damping effect shows itself as reducing the
fluctuations exponentially to a steady state value. The master
equation formalism also allows us to analyze the squeezing for the
thermal state of the relaxed beam (described by the linear
Hamiltonian $H_0$), $\rho(t=0)=e^{-H_0/\theta}/Z$, where
$H_0=a^{\dag}a+1/2$, $\theta=kT/\hbar\omega_0$ and
$Z=Tr(e^{-H_0/\theta})$. Figs. \ref{Fig07}(c) and \ref{Fig07}(d)
show that one need not to start at 0 K to obtain squeezing. As
long as one keeps the noise of the environment sufficiently low,
at higher temperatures of the beam ($kT$ of fractions of the level
spacing $\hbar\omega_0$), although there is no substantial
squeezing initially, at steady state, the amount of ``coordinate"
squeezing is reaching to the
ground state value.  

\section{Conclusion}\label{conclusion}

In this paper we show that squeezed states can be obtained in the
amplitude of the fundamental mode of a nanomechanical oscillator
which has quartic nonlinearity in its effective single particle
quantum mechanical Hamiltonian. The quantum dynamics of the system
is solved numerically both with and without external ac-driving.
For various strength of nonlinearity and driving, the squeezing
dynamics is investigated for both initial ground state and
coherent states of different amplitudes. The terms that lead to
multilevel transitions in the quartic nonlinearity is compared to
similar models and proved advantageous results. It should be noted
that working closer to Euler buckling instability produces larger
squeezing due to large nonlinearity. This is reminiscent of
similar result for quantum optical systems \cite{Wu 1987}.

Finally, we comment on the possibility of observing squeezing in
nanomechanical beams. The obvious way to detect squeezing seems to
be one of the displacement detection methods. There are different
displacement sensing techniques such as optical, magnetomotive and
single-electron transistor (SET) technique. Among these readout
strategies, the SET  --demonstrated by many research groups to be
a very sensitive detector of charge-- enables position detection
for NMO devices in their low energy quantum states. In this
detection method, a charged NMO is capacitively coupled to an SET.
The oscillator's motion induces a change in the charge on the gate
electrode of the SET and the change in the SET's conductance can
be directly monitored. Demonstrations of position detection of
NMO's are developing. The first experiment \cite{Knoebel 2003}
reached the position sensitivity within a factor of 100 of the
quantum limit $\Delta x_{SQL}=\sqrt{\hbar/2m\omega_0}$. A year
after another group \cite{LaHaye 2004} improved this sensitivity
to a factor of 4.3 of the quantum limit set by the Heisenberg
uncertainty principle of a Nanomechanical resonator nearly its
ground state. These demonstrations suggests that beating the
standard quantum limit in the position detection technology is
highly probable in future. Once this is achieved, by reading out
the position each time with a different nonlinearity parameter one
can prove the existence of tunable squeezing. There are also
proposals suggesting the positioning of a NMO inside an
electromagnetic cavity. It is proposed that the interaction of
cavity photons with a NMO in a squeezed state would produce a
nonclassical statistics of photons at the exit. So one can easily
detect squeezing simply by quantum statistical analysis of photons
at the output of the cavity.

In conclusion, in nanomechanical beams described in this paper,
one can remarkably control the quantum squeezing externally just
by controlling the strain provided by a classical compressive
force. However, to control the nonlinearity at the desired values,
one would have to apply compression with extreme delicacy,
$F_c-F_0\sim 10^{-6}F_c$. Controlling the strain to this precision
for sufficient time to identify squeezing may be difficult. Thus,
while observing squeezing will be challenging, the prospect of
exploring tunable quantum squeezing in nanomechanical beams and
the connection to Euler buckling instability are intriguing.

 \end{document}